\documentclass[final]{svjour2}
\usepackage{graphicx}
\usepackage{rotating}
\usepackage{amssymb}
\usepackage{mathptmx}
\usepackage[numbers]{natbib}
\makeatletter

\bibpunct{}{}{,}{s}{}{,}

\begin{document}

\newcommand{\hdblarrow}{H\makebox[0.9ex][l]{$\downdownarrows$}-}
\title{The CUORE cryostat: a 1-ton scale setup for bolometric detectors}

\author{C.~Ligi$^1$ \and C.~Alduino$^2$ \and F.~Alessandria$^3$ \and M.~Biassoni$^{3,4}$ \and C.~Bucci$^5$ \and A.~Caminata$^6$ \and  L.~Canonica$^5$ \and 
L.~Cappelli$^{5,7}$ \and N.I.~Chott$^2$ \and S.~Copello$^6$ \and A.~D’Addabbo$^5$ \and S.~Dell’Oro$^5$ \and A.~Drobizhev$^8$ \and M.A.~Franceschi$^1$ \and
L.~Gladstone$^9$ \and P.~Gorla$^5$ \and T.~Napolitano$^1$ \and A.~Nucciotti$^{3,4}$ \and D.~Orlandi$^5$ \and J.~Ouellet$^9$ \and C.~Pagliarone$^{5,7}$ \and
L.~Pattavina$^5$ \and C.~Rusconi$^3$ \and D.~Santone$^5$ \and V.~Singh$^8$ \and L.~Taffarello$^{10}$ \and F.~Terranova$^{3,4}$ \and S.~Uttaro$^{5,7}$}

\institute{$^1$INFN - Laboratori Nazionali di Frascati, Frascati (Roma) I-00044 - Italy\\
$^2$Department of Physics and Astronomy, University of South Carolina, Columbia, SC 29208 - USA\\
$^3$INFN - Sezione di Milano Bicocca, Milano I-20126 - Italy\\
$^4$Dipartimento di Fisica, Universit\`{a} di Milano-Bicocca, Milano I-20126 - Italy\\
$^5$INFN - Laboratori Nazionali del Gran Sasso, Assergi (L'Aquila) I-67010 - Italy\\
$^6$INFN - Sezione di Genova, Genova I-16146 - Italy\\
$^7$Dipartimento di Ingegneria Civile e Meccanica, Universit\`{a} degli Studi di Cassino e del Lazio Meridionale, Cassino I-03043 - Italy\\
$^8$Department of Physics, University of California, Berkeley, CA 94720 - USA\\
$^9$Massachusetts Institute of Technology, Cambridge, MA 02139 - USA\\
$^{10}$INFN - Sezione di Padova, Padova I-35131 – Italy\\ \\
\email{carlo.ligi@lnf.infn.it}}

\maketitle

\begin{abstract}

The Cryogenic Underground Observatory for Rare Events (CUORE) is a 1-ton scale bolometric experiment whose detector consists of an array of 988 TeO$_2$ crystals arranged in a cylindrical compact structure of 19 towers. This will be the largest bolometric mass ever operated. The experiment will work at a temperature around or below 10~mK. CUORE cryostat consists of a cryogen-free system based on Pulse Tubes and a custom high power Dilution Refrigerator, designed to match these specifications. The cryostat has been commissioned in 2014 at the Gran Sasso National Laboratories (LNGS) and reached  a record temperature of 6~mK on a cubic meter scale. In this paper we present  results of CUORE commissioning runs. Details on the thermal characteristics and cryogenic performances of the system will be also given.

\keywords{Cryogenic intrumentations, Neutrinoless double-beta decay}

\end{abstract}

\section{Introduction}
Neutrinoless Double Beta ($0\nu\beta\beta$) decay can be a source of information about the nature of neutrinos. If this decay exists, it means that neutrinos are Majorana particles (i.e., neutrinos and antineutrinos are equals), and its detection can also provide information about neutrino mass scale and hierarchy. Moreover, in this eventuality, lepton number is not conserved.\\
No confirmed observation of $0\nu\beta\beta$--decay has been officially claimed, so far. The CUORE experiment\cite{artusa} is aimed to search for $0\nu\beta\beta$--decays in $^{130}$TeO$_2$ detectors, using bolometric techniques applied to an array of 988 crystals, arranged in 19 towers. Every crystal is a cube of 5x5x5~cm, 750~g heavy.  The total effective mass of $^{130}$Te is 206~kg.\\
CUORE is located underground, at the Gran Sasso National Laboratories (LNGS) of INFN (Assergi, AQ, Italy). Its detection technology has been successfully tested formerly by CUORICINO, then by CUORE-0 experiments\cite{canonica}.\\
Each CUORE crystal acts both as source and detector. Electron pairs, generated by the $^{130}$Te spontaneous decay, interact with the crystal lattice, releasing energy which is converted in heat, that causes a temperature rise of the crystal itself. In our case this heating is detected by a Neutron Trasmutation Doped (NTD) germanium thermistor, while the crystals need to be operated at temperatures below 10~mK, hence the needs for a dilution refrigerator system. Working at such a low temperature implies a strong vibration damping of the detector, i.e. a very effective mechanical isolation from the ambient. Furthermore, to get a signal with a S/N ratio sufficiently high, the natural background radioactivity around the $0\nu\beta\beta$--decay energy region of interest (2527~keV) must be kept below 10$^{-2}$~counts/keV/kg/year. To match all these requirements, a dedicated cryostat has been designed and built. CUORE is expected to take data for about 5 years.

\section{The CUORE Cryostat}
\begin{figure}
\begin{center}
\includegraphics[
width=11.7cm, height= 8.8cm]
{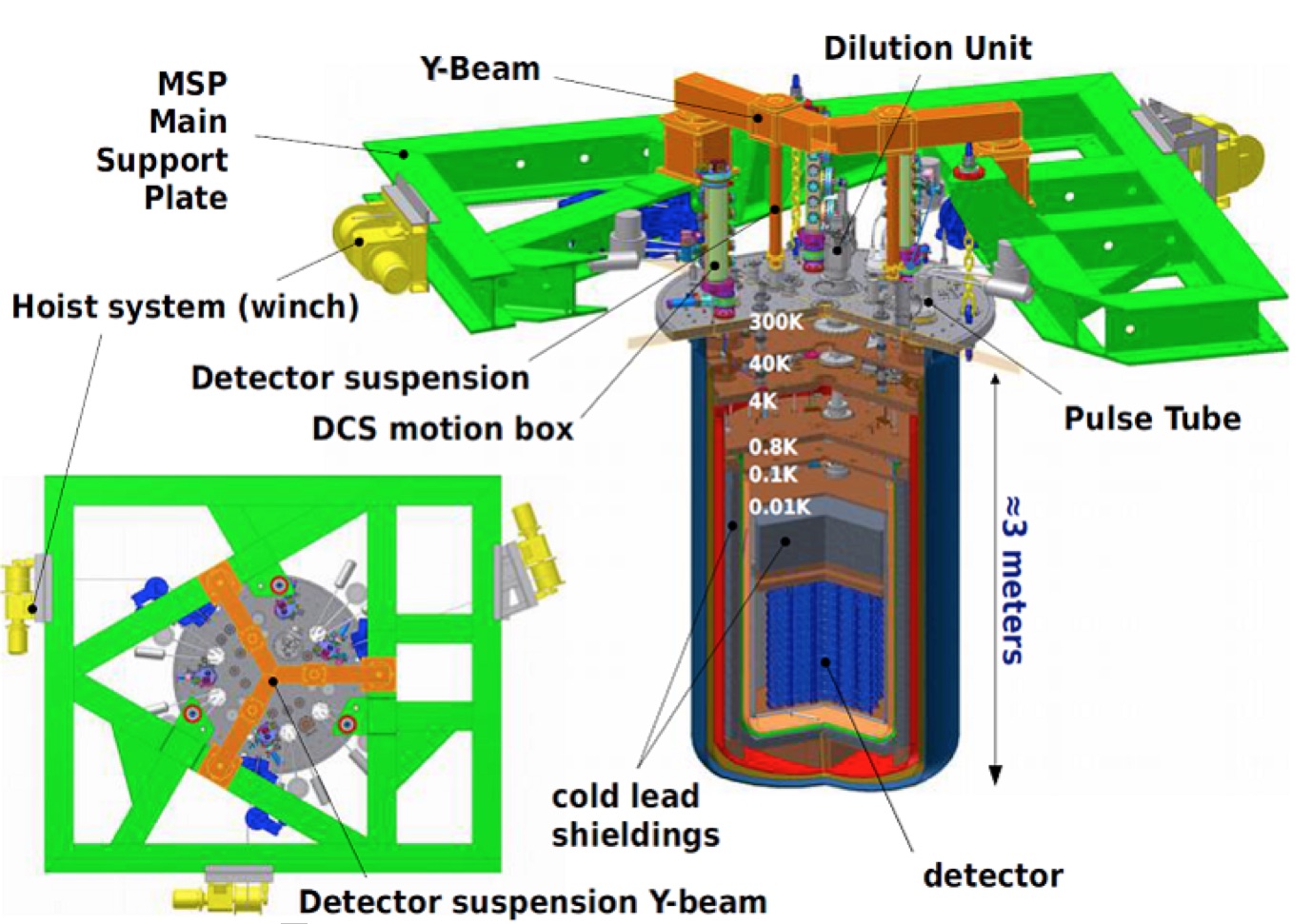}
\end{center}
\caption{Section view of the CUORE Cryostat}
\label{fig1}
\end{figure}

In Fig.~\ref{fig1} a section view of the CUORE cryostat\cite{nucciotti} is shown. It is composed by 6 coaxial shields, kept at 300, 40, 4~K, 600, 50, 10~mK respectively. 300~K and 4~K shields are vacuum tight, and inside the 10~mK shield the detector is placed. The cryostat is suspended from above to a structure called Main Support Plate (MSP). A Y-Beam, isolated from the MSP through {\it Minus-K\textcircled{\check@mathfonts\fontsize\sf@size\z@\math@fontsfalse\selectfont R}} special suspensions, supports the detector. In this way vibration transmission from the cryostat to the crystal is minimized.\\
To get a reasonable radioactive background, CUORE is located underground. To further reduce the residual ambient contamination, the detector is shielded by a set of lead walls, both outside and inside the cryostat. On outside there is a octagonal shape wall, made by lead bricks, that encloses the cryostat on the lateral side and below. Inside the cryostat, just above the detector, is placed a 30~cm thick cylindrical shielding, made of lead sheets, kept at about 50~mK. This is the only shielding wall from the top side. Another cold lead shielding, kept at 4~K, is a 6~cm thick wall that surrounds the 600~mK vessel. Aim of the cold shieldings is to prevent radioactivity contamination from the external vessels (300~K, 40~K and 4~K) and from all the cryostat plates.

\subsection{Pulse Tubes}
Cryostat refrigeration down to 4~K is provided by 5 two-stage Cryomech PT415 Pulse Tubes (PT), with nominal cooling power of 1.5~W at 4.2~K and 40~W at 45~K each. PTs have been  chosen mainly to avoid liquid cryogens refillings, improving the experiment duty cycle. Furthermore, they avoid induced noise from cryogenic liquids boiling.\\
To minimize vibrations from PT rotating motors a {\it remote motor} option has been adopted (accepting a 10\% reduction of the cooling power on both stages), and a rubber cushion is mounted in between of the PTs and the 300~K plate as well. Further damping is provided from the thermal connections between the PT's cold stages and the relative cryostat plates, which are made by flexible thermalizations. PTs have been tested one-by-one in a dedicated cryostat to check their performances and to optimize the He pressure inside the compressor. All of them comply with the Cryomech specification.\\
The cryostat can be kept cold even with only 4 PTs operating, so no warming up is needed in case of one PT's failure or manteinance.

\subsection{Dilution Refrigerator}
Base temperature is reached using a modified DRS-CF3000 continuous-cycle Dilution Refrigerator (DR) made by Leiden Cryogenics. Its nominal cooling power is 5~$\mu$W at 12~mK (3~mW at 120~mK), and the minimum reachable temperature is around 5~mK. It is capable to circulate more than 8~mmoles/s of $^3$He-$^4$He mixture. The Helium mixture flow is controlled by a Gas Handling System which is able to automatically manage the DR circuit valves during all the cooling phases. He mixture flows through two condensing lines in parallel, both for redundancy and to have a flow rate regulation, using either one or two lines at the same time. The condensing lines are thermally connected to two of the five Pulse Tubes, allowing the gas precooling down to 4~K.\\
A peculiarity of this refrigerator is the presence of two spring loaded variable flow impedances, which self-regulate their openings with the incoming flow. In this way, without any external control, it is possible to amplify the flow variations, in order to operate with good efficiency both during the cooling phase (high flow rate) and during the base temperature phase (slow flow rate).\\
The choice to avoid liquid cryogens implies the absence of the 1~K Pot stage, which is needed for the mixture condensation in the traditional DRs. Here, a compressor allows the Helium to circulate in the 4~K stage at a pressure of 2.5~bar, allowing for the condensation of the gas mixture. Lack of the 1~K Pot stage implies that the Still must operate at a relatively high temperature, in the range of 0.8~$\div$~1~K.

\subsection{Fast Cooling System}
The cryostat mass that has  to be cooled at 4~K is about 15~tons. The cooldown using the PTs would take about 5~months. To speed up the process, an apparatus called Fast Cooling System (FCS) was developed. In this way, the cooling time is estimate to shorten to about 2~weeks. The idea is to use forced convection of cold gaseous Helium injected inside the cryostat to cool it. FCS consists of a closed line where Helium is forced to circulate from a compressor. The gas, after being precooled with the outcoming gas from the detector, is then cooled in a small cryostat passing inside a heat exchanger connected with three Gifford-McMahon cryo-coolers with base temperature of about 20~K and cooling capacity of 600~W at 77~K each. After that, the cold gas ($\sim$ 30~K) enters in the CUORE IVC in a tube that drive the gas to flow on the 10~mK vessel. To avoid thermal stresses, Helium flow is regulated to maintain a maximum temperature difference between sample and incoming cold gas, lower than 40~K. FCS commissioning is presently ongoing at LNGS.

\subsection{Detector Calibration System}
Response over the time of the detector at $\gamma$--sources is kept calibrated by an apparatus called Detector Calibration System. Teflon coated copper capsules containing a wire of thoriated tungsten (the radioactive source) are crimped onto Kevlar strings. Several motors drive the strings inside dedicated tubes from the top of the cryostat down to the detectors array level, allowing the calibration of the detector. Strings are thermalized at 4~K before they go into the detector level.

\section{Commissioning Runs}

\subsection{CUORE Cryostat characterization at 4~K}
The CUORE cryostat was first tested in 2013 at 4~K with only 3 PTs and the 300, 40 and 4~K plates and vessels mounted. Two cooldowns were done in this configuration. The main 4~K thermal load, during these runs, was the only 4~K plate and vessel, for a total mass of about 2~tons of copper. During the first run the PT's performances, in terms of base temperature, cooling time and PT behavior during the cooldown, was tested. Steady state temperatures of 3.3~K and 32~K on the 4~K and 40~K stage respectively, were reached. Cooldown from 300 to 4~K took about 1~week, as expected. Some issues on the PT functioning during the first part of the cooldown due to the huge load on the 4~K stage were overcome decreasing (temporary) the Helium inlet pressure inside the PTs. In the second run, the Detector Calibration System was successfully pre-tested at 4~K.

\subsection{Dilution Refrigerator characterization in the test cryostat}
\begin{figure}
\begin{center}
\includegraphics[%
  width=1\linewidth,
  keepaspectratio]{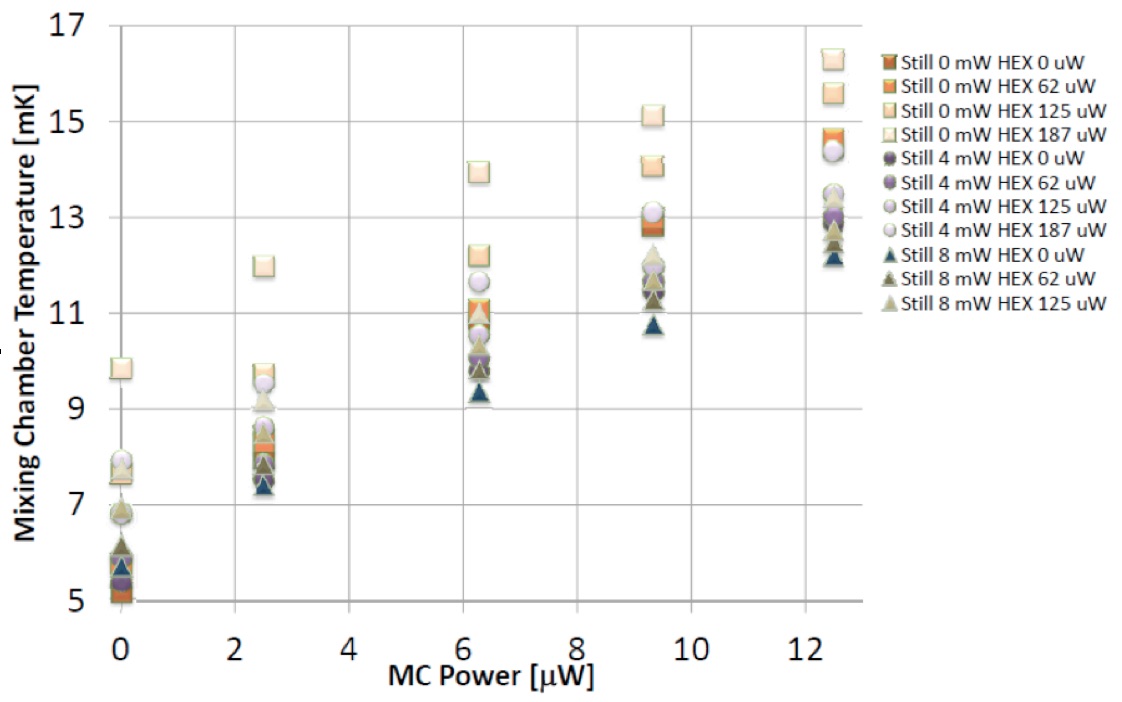}
\end{center}
\caption{Characterization of the dilution refrigerator. MC temperature versus power dissipated on MC, HEX and Still}
\label{fig2}
\end{figure}

The Dilution Refrigerator was first tested in a dedicated test cryostat. Characterization was done by measuring the Mixing Chamber (MC) temperature while injecting power on the cold stages: Still, Heat Exchanger (HEX) and the MC itself. A Cerium Magnesium Nitrate thermometer coupled to a Fixed Point Device was used to monitor the MC temperature.\\
In Fig.~\ref{fig2} the complete data set of the DR characterization is showed. Result were very promising, the measured cooling power being well above the specification. A base temperature of 4.95~mK was achieved, with no power injected. Cooling power at 12~mK was measured to be 9.5~$\mu$W, much higher than the specified value of 5~$\mu$W. As a general comment, putting power on the Still, as expected, produce a decreasing of the MC temperature whenever the MC injected power is greater then 2~$\mu$W (practically always, when the cryostat will host all the components). On the other hand, extra power on the HEX always produced a MC heating.\\
MC temperature stability over one day was of the order of 0.1~mK.

\subsection{Dilution Refrigerator and Wiring characterization in the CUORE cryostat}
In beginning 2014 the DR and the 600, 50 and 10~mK plates and vessels were mounted in the cryostat. These plates are thermally connected to the Still, Heat Exchanger and Mixing Chamber stages of the DR, respectively. Aim of the first base temperature run was the refrigerator characterization verification in the main cryostat.\\
Results were satisfactory. The base temperature was measured to be 5.9~mK. This corresponds to a conductive thermal input to the 10~mK plate of sligthly more than 1~$\mu$W. The DR was capable of cooling the 400~kg copper plate and vessel from 4~K to 10~mK in 12~hours.\\
An issue arised from vibrations coming from the PTs, which caused an heating of the MC plate. These was a low frequency oscillation of the 300~K plate induced by the PTs operation. We fixed it stiffening the plate connection with the Main Support Plate on the horizontal plane. Other issues came from one of the variable flow impedances reliability, which suffered a poor reproducibility of the impedance value during different runs. We are still investigating possible solutions for this malfunctioning. \\
After the DR characterization, another run was dedicated to the wiring thermalization. CUORE wiring consist of 2600 0.1~mm diameter NbTi cables, grouped in 100 ribbons, 13 twisted pair each. Wiring ribbons bring the thermistors and heaters signal from the top plate (300~K) to the MC plate (10~mK) and they can carry heat directly to the bolometers, so their thermalization requires special care. From 300 to 4~K, the ribbons are thermalized by radiation, while below 4~K by conduction, by means of dedicated thermalization clamps connected to the 4~K, 600~mK and 50~mK stages. A dummy mini-tower with 8 crystals was built, mounted and connected to the electronics to check bolometer functionality.\\
During this run, the MC reached a base temperature of 7~mK. Thermistor resistances in the range of the G${\rm \Omega}$ were measured, proving both the goodness of the thermal contact between crystals and MC stage and the effectiveness of wiring thermalization.

\section{Conclusion}
The CUORE cryostat was designed and built. Its commissioning is in its final stage. The tests, done so far, show that the measurements are in good agreement with the expectation and the subsystems work properly. Next step will be the base temperature test with the cold lead shieldings and the FCS mounted, before the detector installation, expected to take place at the beginning of 2016.



\end{document}